# MULTI-VERTEBRAL CT-BASED FE MODELS IMPLEMENTING LINEAR ISOTROPIC POPULATION-BASED MATERIAL PROPERTIES FOR THE INTERVERTEBRAL DISCS CANNOT ACCURATELY PREDICT STRAINS

Chiara Garavelli[1,2], Alessandra Aldieri[1,2], Marco Palanca[2], Luca Patruno[3], Marco Viceconti[1,2]

[1] Medical Technology Lab, IRCCS Istituto Ortopedico Rizzoli, Bologna (IT)

[2] Department of Industrial Engineering, Alma Mater Studiorum - University of Bologna (IT)

[3] DICAM, University of Bologna, Bologna, Italy

**ORCID IDs**

| | |
|---|---|
| Chiara Garavelli | https://orcid.org/0000-0002-6921-0730 |
| Alessandra Aldieri | https://orcid.org/0000-0002-2397-3353 |
| Marco Palanca | https://orcid.org/0000-0002-1231-2728 |
| Luca Patruno | https://orcid.org/0000-0003-4973-207X |
| Marco Viceconti | https://orcid.org/0000-0002-2293-1530 |

# MULTI-VERTEBRAL CT-BASED FE MODELS IMPLEMENTING LINEAR ISOTROPIC POPULATION-BASED MATERIAL PROPERTIES FOR THE INTERVERTEBRAL DISCS CANNOT ACCURATELY PREDICT STRAINS

**ABSTRACT**

Vertebral fractures cannot currently be accurately predicted in clinics, where the adopted scores have limitations in identifying subjects at risk. In this context, computed tomography (CT)-based finite element (FE) models might improve fracture risk prediction. Many proposed FE models consider single vertebrae only, thereby neglecting the role of the intervertebral discs in load transmission and distribution across vertebrae. Multi-vertebrae models would allow the inclusion of more physiological boundary conditions in the simulation thanks to the discs' inclusion. Nevertheless, while CT allows material properties to be assigned to the vertebrae according to the Hounsfield Units, no information about the discs' mechanical properties is provided. Hence, the aim of this study was to validate a CT-based multi-vertebrae FE model where linear isotropic material properties were assigned to the discs against experimental data. The CT-based FE model of a multi-vertebrae specimen was built assigning populations values retrieved from the literature to the discs Young's modulus and applying boundary conditions coherently with the experiments performed on the specimen. Computational strains on the vertebrae surfaces were compared against the experimental strains coming from digital image correlation measurements. The strains on the vertebrae surfaces increased following the increase of the discs Young's modulus, with no changes in their local distributions. Although Young's modulus values around 25-30 MPa yielded comparable orders of magnitude between numerical and experimental strains, strains local distribution differed substantially. In conclusion, a different modelling approach should be adopted for the discs in CT-based multi-vertebrae FE models to achieve acceptable accuracy.

**KEYWORDS**

**INTRODUCTION**

Vertebral fracture represents a very severe event, which, regardless of the pathology associated with its occurrence, increases morbidity, mortality and decreases the quality of life in already fragile subjects [1]. In addition, it also has economic implications for the healthcare systems [2]. According to the World Health Organization (WHO), by 2030, the number of people older than 60 will increase by 34% compared to 2020 [3]. Due to risk factors such as increased propensity to fall and deteriorated mechanical properties of the bone tissue, the elderly present a markedly increased bone fracture incidence, mainly at the hip, wrist, and spine [4]. At the same time, the WHO has also reported that almost 20 million new cancer cases are diagnosed every year [5]. In fact, the improved therapeutic protocols for cancer patients have significantly increased their average survival [6]. In return, an increasing incidence of bone metastases has been observed, primarily located in the spine [7], which compromises the physiological mechanical competence of the vertebrae [8], [9]. The spine is already per se one of the most fracture-prone anatomical sites [10], and pathological vertebrae are exposed to an even higher risk of fracture [11]. Therefore, an accurate prediction of the fracture risk turns out to be pivotal so that appropriate preventive interventions are taken.

The available clinical gold standards for bone fracture prediction have shown limited accuracy in stratifying subjects at high risk of experiencing a fracture from subjects who are not. The widely used T-score, for example, was proved to suffer from poor specificity and sensitivity in predicting hip and vertebral fractures connected to osteoporosis [12]. Similarly, the Spinal Instability Neoplastic Score (SINS), adopted in clinics for referring patients to surgical/orthopaedic consultation in case of metastatic vertebrae, suffers from a certain degree of subjectivity in the final clinical decision [13] and lacks sufficiently good specificity [14]. Nevertheless, an accurate prediction of the risk of fracture would be crucial, especially in metastatic subjects, as surgeons could decide whether to surgically intervene on subjects already weakened by primary cancer treatment.

In recent years, Computed Tomography (CT)-based digital twins development has highlighted the potentiality of such *in silico* methodologies in this context. Several studies have demonstrated the possibility of adopting *in silico* tools to provide information about bone resistance non-invasively [15]–[17]. For some anatomical sites, such as the femur, *in silico* methodologies have successfully outperformed the gold standard for predicting hip fracture [18], [19]. On the contrary, due to its complexity, the fracture prediction at the spine *in silico* is not as straightforward.

The spine comprises multiple vertebrae with intervertebral discs (IVDs) interposed, which affect the load transmission and distribution. Consequently, vertebral fracture risk prediction *in silico* based on clinical images (e.g., CT) could be performed by adopting two mutually exclusive strategies. In the first one, the vertebra fracture risk is predicted by only modelling one vertebra and neglecting the IDVs [20]–[22]. This approach intrinsically poses some limitations since it does not allow to impose physiological loading conditions on the vertebral bodies and does not consider to what extent IVD degeneration might impact load transmission to the adjacent vertebrae [23]–[25]. On the other hand, the second modelling strategy includes the IVDs in the model by modelling at least one entire vertebral unit to account for the real complexity of the spine. In principle, this would require the inclusion of a comprehensive description of the IVD through magnetic resonance imaging (MRI) as an additional input to the model besides CT, as it would allow for the quantification and integration of the patient-specific IVDs structure and composition [26], [27], [28]. Although this approach would increase the accuracy of the model, allowing to reproduce a more realistic load transmission among vertebrae, it brings its own set of complications (e.g., higher costs and higher clinical workload), preventing its implementation in clinical practice. In fact, the development of a digital twin for vertebral fracture risk prediction based only on the subject's CT would be much more clinically feasible. The CT would allow reconstruction of the geometry and the assignment of patient-specific Hounsfield Units-based material properties to the vertebral bone tissue. However, no information would be provided regarding the disc, forcing to model it with population material properties coming from the literature.

In this light, the aim of this work was to assess the accuracy of a multi-vertebrae FE model uniquely based on CT, where the IVDs were included and modelled in the simplest possible way with homogenous and linear isotropic material properties. A patient-specific FE model of a spine segment with evidence of metastatic disease was developed from CT images of a cadaveric sample. A range of population-based Young's modulus values from the literature [29] were assigned to the IVDs. The vertebral strains obtained from the model were compared to experimental data acquired through Digital Image Correlation (DIC) [30] on the same sample, and the accuracy of the proposed modelling strategy assessed.

## MATERIALS AND METHODS

### Mechanical Testing

The procedures for conducting the mechanical testing have been thoroughly explained in a prior publication [31] and will be briefly summarized here. Experimental tests were performed on a thoracolumbar cadaveric specimen obtained from an ethically approved donation program (Anatomic Gift Foundation, Inc.). The sample included four vertebrae and the interposed three IVDs, from T10 to L1. One of the two central vertebrae (T12) showed signs of lytic metastatic lesions (Fig.1A). Half of the most cranial and half of the most caudal vertebrae were embedded in polymethyl-methacrylate (PMMA) cement (Fig.1B). After the preparation of the specimen, diagnostic images were acquired with a spiral CT (AquilionOne, Toshiba, Japan), to be subsequently used for the FE model development. Scans were performed at 120 kVp with 200 mA tube current, 0.24 mm × 0.24 mm pixel size and 1 mm slice thickness. Afterwards, the specimen was sprayed with white water-based paint to generate a unique speckle pattern recognisable by a three-dimensional DIC 4-camera system (Aramis Adjustable 12M, GOM, Braunschweig, Germany, with 12mPixels cameras and 75 mm metrology-standard lenses). Flat circular markers were glued on the aluminium pots of the test machine to track the displacement of the superior and inferior pots (Fig.1C).

The mechanical tests were performed using a uniaxial testing machine (Instron 8500 controller with Instron 25 kN load cell, Instron, UK) in displacement-controlled modality. The load application point was shifted forward by 10% of the anteroposterior dimension of the central IVD to impose an anterior flexion [32]. The use of a ball-joint and low friction bearings allowed the top pot to be free to rotate and translate, while bottom one was totally constrained. The load magnitude was identified as the load able to induce minimum principal strains typically associated to physiologic loads [33] (i.e. in the order of 2500/ 3000 microstrain) on the anterior surface of the control vertebra. The experimental reaction force was recorded. Further details can be found in [31]. During the test, DIC images of the two central vertebrae free from the cement and of the two pots were acquired at 25 Hz (Fig.1D). DIC measured the displacements on the visible vertebral surfaces. Systematic and random errors in term of displacements were found around 10 μm and 25 μm, respectively. Maximum and minimum principal strains on the vertebrae surface were then calculated by derivation of the nodal displacements, as described later in the Metrics paragraph. The DIC spatial resolution was about 2 mm.

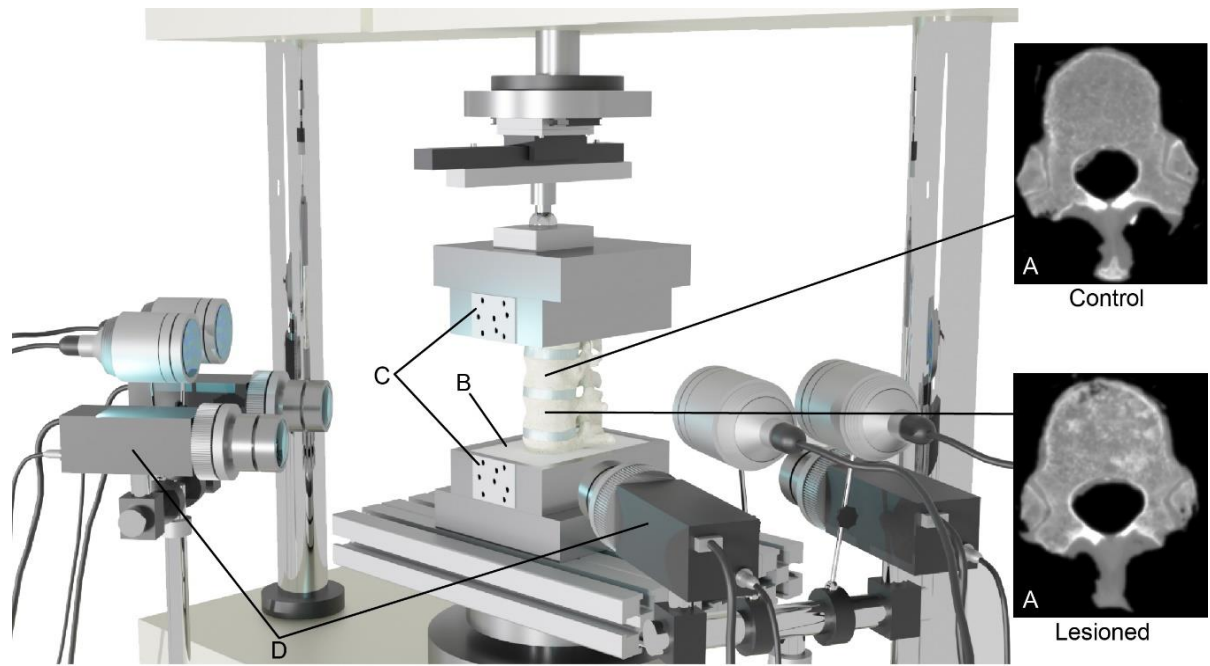

**Fig.1: Experimental test set-up.** The specimen with one control (A, top) and one metastatic (A, bottom) is placed into the testing machine (B). Pots displacements are tracked thanks to glued markers (C). The flexion test is recorded by a three-dimensional DIC system (D).

**Finite Element Analysis**

The steps required to develop a CT-based homogenised FE model have already been presented in detail in our previous work [34] and will be here briefly summarised. Firstly, CT images of the specimen were segmented to extract the geometry of the vertebral bodies, IVDs, and pots separately. A threshold method based on Hounsfield Units (HU) values was used to pick out bone tissue and PMMA cement (200÷3000 HU for bone and 1100÷1800 HU for PMMA), followed by manual editing (Mimics 25.0, Materialise NV, Leuven, Belgium). The IVDs profiles were segmented completely manually: lateral contour was sufficiently well visible from the CT scan, while adjacent vertebrae endplates were used as upper and lower bounds. Secondly, subtraction Boolean operations were performed to glue vertebral bodies with adjacent intervertebral discs to obtain a unique volume object (SpaceClaim V19.3, Ansys Inc., Canonsburg, PA). Eventually, posterior processes were removed, considering that in the flexion test, the posterior ligaments were activated only if the physiological range is exceeded (at this spinal level, the neutral zone and the range of motion for each FSU are around 0.6° ± 0.1° and 3.5° ± 0.8° respectively [35]), which instead was guaranteed in the presented test-case (total flexion: 2.8°).

A ten-node tetrahedral structural solid mesh was generated following a sensitivity analysis, imposing a maximum edge length equal to 2 mm (ICEM CFD V19.3, Ansys Inc.). Subsequently, material properties were mapped on each element (Bonemat® V3.1, Istituto Ortopedico Rizzoli, Bologna, Italy) [36]. Bone tissue was modelled as a heterogeneous linear elastic isotropic material. Specifically, the voxel HU values were converted to equivalent values of volumetric bone mineral density through a phantom-based calibration and the application of an additional calibration correction [37]. A density-elasticity relationship [24] was adopted to assign heterogeneous elastic properties over space to the vertebral elements. A Poisson's coefficient (ν) equal to 0.3 was assigned to all bone elements. The IVDs were modelled as isotropic and homogeneous linear elastic materials, all characterised by the same elastic modulus ($E_{disc}$). Initially, $E_{disc}$ was set to the value that allowed it to fit the experimentally recorded reaction force, as described in [34]. Subsequently, the analysis was extended to the entire range covered by the literature, ranging from 4 MPa to 50 MPa [29], [39]–[41]. A Poisson's coefficient (ν) equal to 0.1 was adopted [42]. Linear elastic material properties were also assigned to the PMMA pot, with ν=0.3 and an elastic modulus ($E_{PMMA}$) equal to 3 GPa (obtained from in-house experimental tests on cement samples).

Boundary conditions were assigned to the FE model to replicate the experimental test conditions. The metal and PMMA cement pots were assumed to be rigidly connected so that only the second ones were modelled. The positions of the superior metal pot at the initial and final steps of the experimental test acquired through the DIC, were processed using a single value decomposition (SVD) algorithm to extract the rotation components and translation of the rigid pot's motions. The obtained values were imposed on all the external nodes of the FE model superior PMMA pot using a multi-point constraint. The inferior PMMA pot was fixed.

Simulations were solved in the FE software with the preconditioned conjugate gradient solver, using parallel distributed memory over 6 cores with 64 GB of RAM (Intel(R) Xeon(R) E-2276G CPU 3.80GHz).

Maximum and minimum principal strains were calculated on the surface of the vertebrae by derivation of the numerical nodal displacements, as described later in the next paragraph.

**Metrics**

The DIC and FE reference systems had to be aligned prior to the comparison between experimental and numerical outcomes. This was done adopting a procedure based on surface registration (Mimics 25.0) and SVD (Matlab® v2020, MathWorks, Natick, Massachusetts, US), as described in details elsewhere [34].

Subsequently, DIC vertebral displacements were interpolated onto the locations of the superficial nodes of the FE model. To do this, an inverse distance weighting algorithm based on the Euclidean norm was used, with power equal to 2 and threshold radius set to 1 mm accounting for the surface registration error. The interpolation allowed to perform pointwise comparison between the experimental and the numerical outcomes (Matlab® v2020), for all the different $E_{disc}$ values applied.

In order to check if the kinematics observed in the experiments was correctly reproduced by the model, DIC and numerical displacements agreement was assessed point wise. To evaluate this, linear regression coefficients, determination coefficient ($R^2$) and root mean squared error normalized by the maximum measured value (%RMSE) were computed between DIC and FE local displacements.

Experimental and numerical strains on the vertebral surfaces were obtained through derivation of the displacement fields on the superficial nodes of the vertebrae. More in detail, the strains were calculated by defining linear triangular elements as the faces of the mesh tetrahedrons lying on the vertebral surface. Maximum and minimum principal strains were considered. $R^2$, RMSE, and %RMSE between experimental and FE strain fields were evaluated in order to identify, if possible, the $E_{disc}$ value providing the best match between the two fields. In addition, also percentual differences between FE model and DIC were computed on strains. Kolmogorov-Smirnov test was adopted to compare DIC and FE principal strains distributions.

Each vertebral surface was also divided into three Regions of Interest (RoI), corresponding to the central, the right and the left portion of the external surface, where the experimental and numerical were averaged and compared.

**RESULTS**

Numerical and experimental displacements were compared and showed a %RMSE lower than 9% (Fig. 1s in the Electronic Supplementary Material).

A $E_{disc}$ equal to 4.15 MPa allowed fitting of the experimental reaction force. Nevertheless, experimental and numerical principal strains differed of more than one order of magnitude, with the model underestimating the experimental values (RMSE > 85%). As $E_{disc}$ increased from 4.15 to 50 MPa, the numerical strains increased without changes in the numerical strains local distribution, which resulted in a considerable disagreement (p<0.001) between the numerical and experimental strains distributions (Fig. 2).

The values of $R^2$, RMSE, and RMSE% between experimental and FE local strain measurements for all the $E_{disc}$ considered are reported in Table 1s in the Electronic Supplementary Material for both the control and the lesioned vertebra. For the $E_{disc}$ assignment (25 MPa) providing the best fitting of the experimental strain field, RMSE% higher than 30% and $R^2$ lower than 0.55 were found. Average percentual strain differences between FE and DIC settled to 20% and 65% for minimum and maximum principal strains respectively on the control vertebra, and to 35% and 55% respectively on the metastatic vertebra. Errors did not exceed 90% in any case.

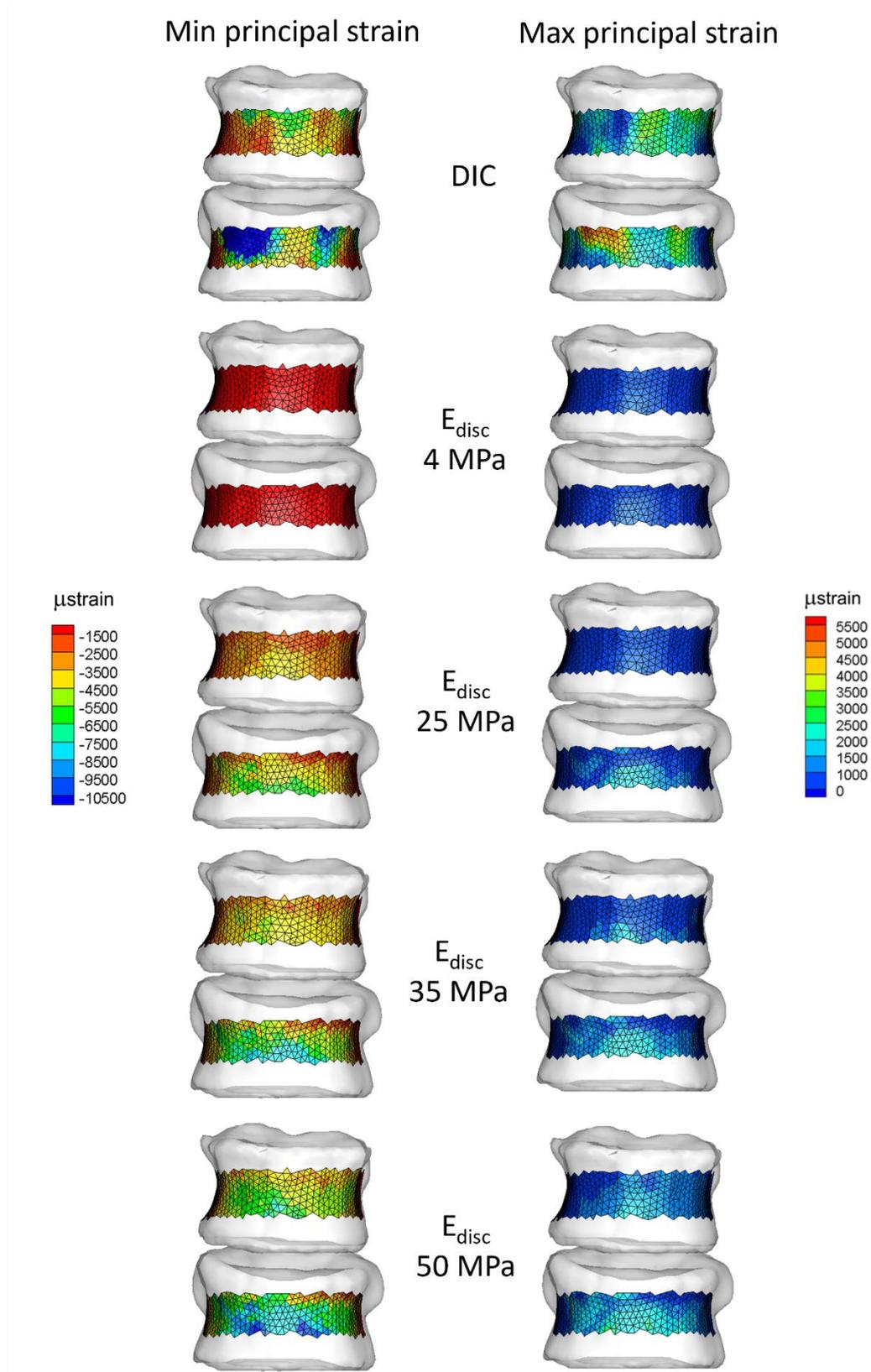

**Fig.2: Contour plots of the computational principal strains at different $E_{disc}$.** Comparison between experimental (first row) and computational (second to final row) principal strains over the vertebrae surface, for the different $E_{disc}$ implemented.

Fig. 3 and Fig. 4 present the comparison between the average values of the maximum and minimum principal strains in all the three RoIs. As visible, the twelve times stiffening of the IVDs made the strains intensity on the vertebrae about eight times in compression and nine times in traction (Fig.3). Moreover, $E_{disc}$ equal to 25 MPa was found to minimise the %RMSE computed on the minimum principal strains (Fig.4) for both control and lesioned vertebra. None of the implemented $E_{disc}$ values turned out to be able to minimise the %RMSE computed on the maximum principal strains.

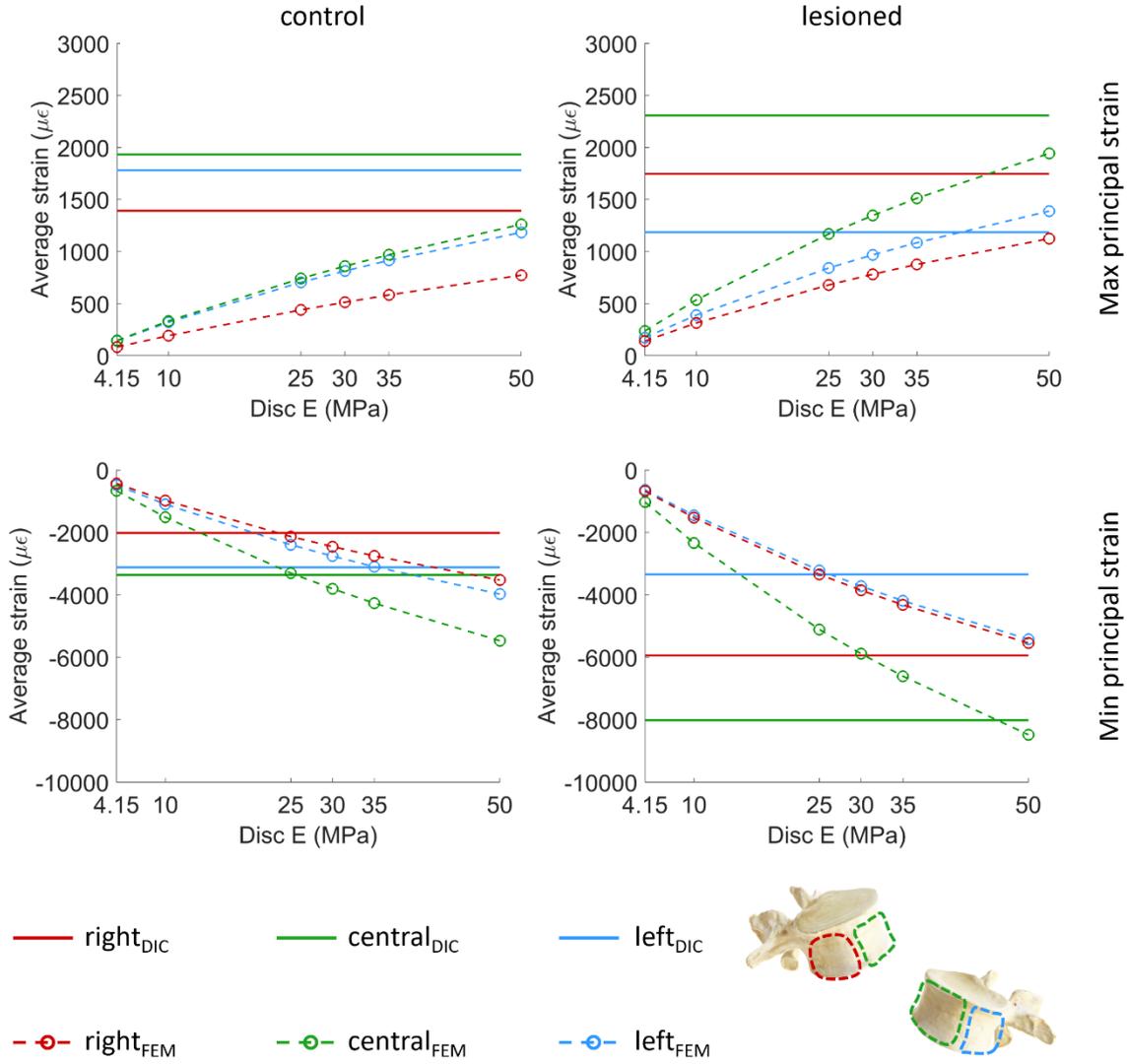

**Fig.3: Strains dependence to the elastic modulus.** Comparison between numerical (dotted lines) and experimental (solid line) averaged on the RoI for the $E_{disc}$ values tested.

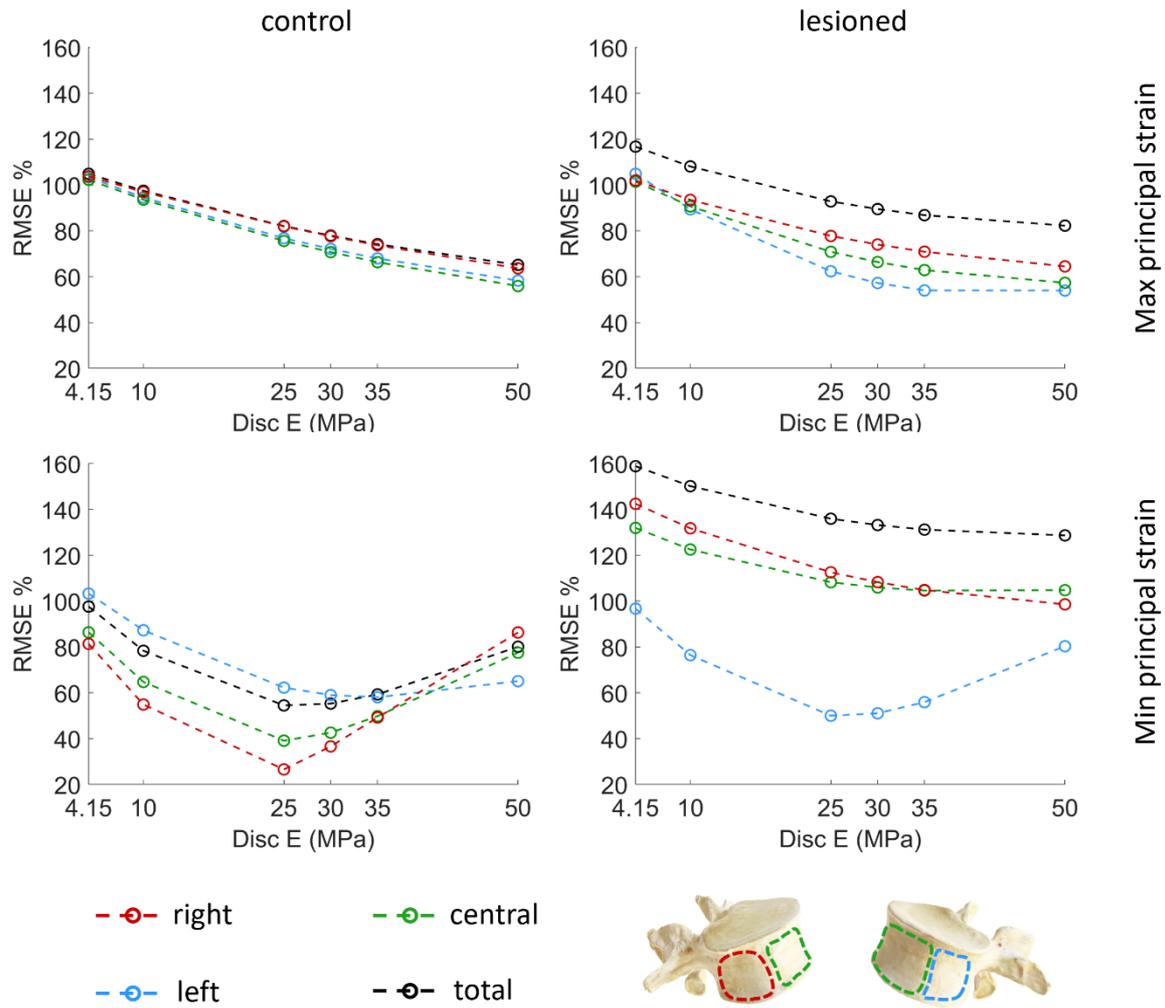

**Fig.4: Strains error dependence to the elastic modulus.** %RMSE computed between numerical and experimental strains for each RoI (red, blue, green dotted lines) and for the whole vertebra (black dotted line) at the $E_{disc}$ values tested.

## DISCUSSION

The purpose of the presented study was to assess whether a CT-based multi-vertebrae FE modelling framework turned out to be accurate and reliable enough to be employed in clinical practice. If so, such framework could represent a valid tool to support and improve the currently adopted tools for vertebral fracture risk prediction. A FE model of the human spine was developed uniquely from CT information. The simplest assumption possible was made regarding the IVDs material properties, adopting a homogeneous linear isotropic behaviour. The FE outcomes were eventually compared to the corresponding DIC experimental results to assess the model's accuracy [31]. CT-based FE models for risk of fracture assessment in vertebrae are usually developed considering only one vertebra as an isolated body [43] [44]. However, this strongly prevents the application of physiological loading conditions, such as those provided by the IVDs [45], as well as the occurrence of high strain concentration close to the endplates, where fracture often originates [46]. The IVDs play a crucial role in load transmission among vertebrae [24]. Since IVDs can deeply influence the vertebrae deformation state [23], *in silico* vertebral fracture prediction without IVD can turn out to be jeopardised. Thus, developing reliable predictive models would require the inclusion of at least one functional spine unit. On the other hand, accurate and patient-specific modelling of the IVD mechanical behaviour requires information from the MRI imaging technique: this would imply the need for two distinct techniques, MRI and CT, to build a multi-vertebrae FE model for predicting the fracture risk at the spine, making its application in clinical practice more difficult. The main objective of the presented study was therefore to assess whether a CT-based multi-vertebrae FE model could be a sufficiently accurate predictor of vertebrae deformation state. The crucial role played by the disc was also highlighted in the work of Hussein and coworkers [47], where experimental compression tests were carried out loading the vertebra through the IVDs and employing digital volume correlation (DVC), the volumetric extension of the DIC, to measure the displacements. That allowed to develop single vertebra FE models where both the actual displacements detected on the superior vertebral endplate through DVC and a uniform compression were applied. Considerable differences were observed by comparing numerical and experimental vertebral displacements in the two cases (DVC-derived BCs: $R^2 = 0.66$, percentage differences = 9–48%; uniform idealized BCs: $R^2 = 0.02$, percentage differences = 20–240%;).

The here presented findings showed that variations in the order of magnitude of the numerical deformations could be observed consequently to $E_{disc}$ increment, although without any apparent changes in the deformations distribution over the vertebral surfaces. These results were in agreement with the ones presented by Yang et al. [29], where the disc effective spatial stress distribution was analysed.

To the best of the authors' knowledge, there is only another study performing full-field validation on the vertebral surface strains field [48], where only a single vertebra was modelled. There, a good accuracy was found in the longitudinal strains prediction, with 50% of the nodes showing a 12.5% accuracy, while a lower agreement with experimental strains was found in the circumferential strains, with median error of 39% and 10% of the nodes over 100% error. Herein, if the best case ($E_{disc}$ = 25 MPa) is considered, 20% and 65% average differences were obtained for minimum and maximum principal strains respectively on the control vertebra, while 35% and 55% average differences for minimum and maximum principal strains were found on the metastatic vertebra. None of the nodes showed errors higher than 90%. Imai et al. [15] also worked on single vertebrae, validating the predicted minimum principal strains against experiments where four rosette strain gauges were applied on the vertebral surface. Good agreement was found between numerical and computational values (slope = 0.93, intercept = 74 µε, $R^2 = 0.84$). Also Clouthier et al. [49] worked on spine segments FE models, but they performed validation on the ultimate load values, not available here in light of the conservative design of the study.

Other studies also tried to validate single vertebral FE models through the comparison of experimental failure locations versus simulated strains contour [50] or simulated damage accumulation [51]. Herein, a consistency between the highest strains regions according to the predictions and the experiments could not be found. On the other hand, the vertebra experiencing the highest strains degree was consistent to DIC results.

This study presents some limitations which are described in the following. A first critical consideration involves the mechanical properties assigned to the intervertebral discs, which, as claimed from the beginning, were modelled adopting the simplest approach possible. Further investigation could be carried out to slightly increment the complexity of the adopted constitutive laws and assign different population-based properties to each IVD component (i.e., annulus and nucleus). Additionally, more sophisticated FE models would be required to explore the influence of posterior ligaments and facet joints, which were omitted in this work due to their limited role in the considered loading conditions. Still, a CT-based FE model could not include subject-specific information about soft tissues composition and real mechanical behaviour. Moreover, the IVD elastic modulus was shown to be correlated to their degenerative status in [27], heavily affecting how the load is transferred across the spine and, consequently, the vertebral body deformations. It is then still to be determined whether a multi-vertebrae FE model developed uniquely from CT could provide sufficiently accurate insights into the vertebral fracture risk. The second area of limitations concerns the boundary conditions assignment. In fact, it was assumed that no relative motion occurred between the cranial potted vertebra and the cement it was potted in as well as between the cement and the metal pot in which it was inserted and locked. However, this assumption could be judged reasonable due to the considerably greater stiffness of the metal pots compared to the specimen, also considering the secure connection ensured by the screws. The third point is related to the material properties assigned to the vertebral bone. The same constitutive laws were employed for all the vertebrae analysed in the different studies, without considering the vertebrae' health status. While this choice may have introduced some inaccuracies, it was aligned with the findings of other authors who observed similarities in the mechanical properties of trabecular bone with and without lesions [52], [53]. This assumption holds when metastatic lesions can be characterized as low-density bone tissue, as in the case of lytic lesions, while it might not be accurate enough for blastic lesions. The specimen used for this study showed mixed metastases, with a limited occurrence of blastic lesions. Eventually, it should also be noticed that the developed method was presented on one specimen only, considering the methodological nature of the work.

In summary, the obtained findings suggest that CT-based multi-vertebrae FE models including subject-specific multiple vertebrae and IVDs modelled with homogenous linear isotropic material properties based on population values were not able to accurately predict strain distribution on the vertebral surfaces.

**ACKNOWLEDGEMENTS**

The authors would like to thank Luigi Lena for the illustrations.

**CONFLICT OF INTEREST**

This study was supported by the European Commission through the H2020 project "In Silico World: Lowering barriers to ubiquitous adoption of In Silico Trials" (topic SC1-DTH-06-2020, grant ID 101016503). Some of the computational aspects were optimised with the support of the European Commission through the H2020 project "CompBioMed2: A Centre of Excellence in Computational Biomedicine" (topic INFRAEDI-02-2018, grant ID 823712). The study was also partially funded by the AOSpine Discovery and Innovation Awards (AOSDIA 2019_063_TUM_Palanca), Marie Skłodowska-Curie Individual Fellowship (MetaSpine, MSCA-IF-EF-ST, 832430/2018)

The authors declare that they do not have any financial or personal relationships with other people or organisations that could have inappropriately influenced this study.

**OPEN ACCESS DATA**

The following Open Access Data are linked to this manuscript:

DOI 10.6092/unibo/amsacta/7442

# SUPPLEMENTARY MATERIAL

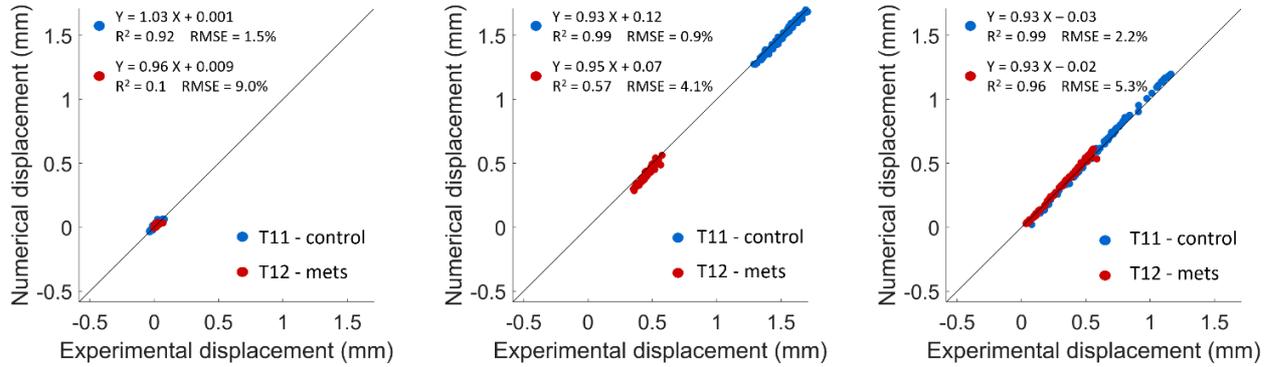

**Fig.1s**: Comparison between the displacements intensity measured using DIC and the ones predicted by the FE model on the vertebral surfaces. Linear regression, coefficient of determination and root mean squared error are also reported.

|  | | control | | | | | | | | lesioned | | | | | | | |
|---|---|---|---|---|---|---|---|---|---|---|---|---|---|---|---|---|---|
|  | $E_{disc}$ (MPa) | left | | central | | right | | total | | left | | central | | right | | total | |
|  |  | $\varepsilon_{MIN}$ | $\varepsilon_{MAX}$ | $\varepsilon_{MIN}$ | $\varepsilon_{MAX}$ | $\varepsilon_{MIN}$ | $\varepsilon_{MAX}$ | $\varepsilon_{MIN}$ | $\varepsilon_{MAX}$ | $\varepsilon_{MIN}$ | $\varepsilon_{MAX}$ | $\varepsilon_{MIN}$ | $\varepsilon_{MAX}$ | $\varepsilon_{MIN}$ | $\varepsilon_{MAX}$ | $\varepsilon_{MIN}$ | $\varepsilon_{MAX}$ |
| **RMSE ($\mu\varepsilon$)** | 4.15 | 3439 | 2166 | 3464 | 2067 | 1945 | 1472 | 2955 | 2030 | 3789 | 1396 | 8023 | 2307 | 8333 | 1944 | 7515 | 2096 |
|  | 10 | 2931 | 2005 | 2762 | 1903 | 1408 | 1366 | 2407 | 1890 | 3202 | 1232 | 7084 | 2064 | 7532 | 1767 | 6797 | 1915 |
|  | 25 | 2017 | 1666 | 1600 | 1552 | 528 | 1128 | 1477 | 1589 | 2201 | 921 | 5413 | 1584 | 5993 | 1419 | 5463 | 1562 |
|  | 30 | 1833 | 1572 | 1466 | 1453 | 562 | 1058 | 1342 | 1504 | 2039 | 846 | 5068 | 1463 | 5616 | 1330 | 5157 | 1471 |
|  | 35 | 1709 | 1485 | 1471 | 1363 | 741 | 992 | 1301 | 1426 | 1966 | 784 | 4819 | 1356 | 5297 | 1252 | 4910 | 1393 |
|  | 50 | 1660 | 1265 | 1988 | 1130 | 1437 | 817 | 1584 | 1223 | 2162 | 666 | 4570 | 1137 | 4642 | 1073 | 4480 | 1216 |
| **RMSE%** | 4.15 | 105 | 101 | 86 | 100 | 84 | 106 | 100 | 108 | 96 | 104 | 132 | 102 | 141 | 102 | 157 | 116 |
|  | 10 | 89 | 93 | 65 | 92 | 53 | 96 | 79 | 96 | 77 | 89 | 122 | 91 | 132 | 94 | 149 | 109 |
|  | 25 | 65 | 75 | 39 | 74 | 28 | 84 | 53 | 84 | 49 | 63 | 108 | 71 | 112 | 78 | 135 | 93 |
|  | 30 | 62 | 71 | 43 | 70 | 35 | 80 | 56 | 80 | 52 | 58 | 104 | 66 | 108 | 74 | 128 | 90 |
|  | 35 | 56 | 70 | 50 | 69 | 47 | 72 | 62 | 72 | 57 | 54 | 104 | 64 | 104 | 71 | 96 | 87 |
|  | 50 | 68 | 60 | 77 | 59 | 85 | 62 | 81 | 65 | 80 | 54 | 104 | 58 | 98 | 65 | 77 | 84 |
| **$R^2$** | 4.15 | <0.1 | <0.1 | 0.55 | <0.1 | 0.46 | 0.21 | 0.15 | 0.10 | 0.20 | 0.14 | <0.1 | <0.1 | 0.24 | <0.1 | 0.11 | <0.1 |
|  | 10 | <0.1 | <0.1 | 0.54 | <0.1 | 0.46 | 0.20 | 0.15 | 0.10 | 0.20 | 0.14 | <0.1 | <0.1 | 0.24 | <0.1 | 0.11 | <0.1 |
|  | 25 | <0.1 | <0.1 | 0.54 | <0.1 | 0.46 | 0.17 | 0.15 | 0.10 | 0.20 | 0.13 | <0.1 | <0.1 | 0.25 | <0.1 | 0.11 | <0.1 |
|  | 30 | <0.1 | <0.1 | 0.53 | <0.1 | 0.47 | 0.17 | 0.14 | 0.10 | 0.20 | 0.13 | <0.1 | <0.1 | 0.25 | <0.1 | 0.11 | 0.10 |
|  | 35 | <0.1 | <0.1 | 0.53 | <0.1 | 0.48 | 0.16 | 0.14 | 0.10 | 0.20 | 0.13 | <0.1 | <0.1 | 0.26 | <0.1 | 0.11 | 0.10 |
|  | 50 | <0.1 | <0.1 | 0.53 | <0.1 | 0.48 | 0.14 | 0.14 | 0.10 | 0.20 | 0.12 | <0.1 | <0.1 | 0.26 | <0.1 | 0.11 | 0.11 |

**Table 1s**: Correlation indexes between predicted and experimental local displacements. RMSE, RMSE%, and $R^2$ between experimental and FE local strain measurements for all the $E_{disc}$ considered were reported, for both the control and the lesioned vertebra, considering each RoI as well as the total analysed lateral surface.